\documentclass[a4paper,11pt]{article}
\usepackage{jheppub}
\usepackage{lineno}
\usepackage{amsmath,mathrsfs,multicol,multirow,array}
\usepackage{booktabs}
\usepackage{tabularx}
\usepackage{adjustbox}
\usepackage{comment}
\usepackage{mathrsfs}
\usepackage[utf8]{inputenc}
\nolinenumbers

\allowdisplaybreaks

\title{\boldmath Quantum Fisher Information Revealing Parameter Sensitivity in Long-Baseline Neutrino Experiments}

\author[a,b]{Bhavna Yadav}
\author[a,b]{Amir Subba}
\author[a,b]{Yu Shi}

\affiliation[a]{Wilczek Quantum Center, Shanghai Institute for Advanced Studies, Shanghai 201315, China}
\affiliation[b]{University of Science and Technology of China, Hefei 230026, China}
\emailAdd{bhavnayadav04@ustc.edu.cn}
\emailAdd{amirsubba@ustc.edu.cn}
\emailAdd{yu\_shi@ustc.edu.cn}

\abstract{Determination of the leptonic CP-violating phase $\delta_{\rm CP}$, the atmospheric mixing angle $\theta_{23}$, and the mass-squared difference $\Delta m_{31}^2$ constitutes a primary objective of current and next-generation long-baseline neutrino experiments. We employ Quantum Fisher Information (QFI) to quantify the maximum information that the neutrino quantum state contains about these oscillation parameters, treating the neutrino as an evolving pure quantum state. Computing the QFI as a function of the baseline-to-energy ratio $L/E$ for benchmark parameter sets from NuFit-6.0, we find distinct sensitivity hierarchies and $L/E$-dependent structures. Specifically, $\delta_{\rm CP}$ and $\theta_{23}$ exhibit bimodal QFI profiles with peaks around $L/E \sim 500$ and $1500~\mathrm{km/GeV}$, reaching $F_Q(\delta_{\rm CP}) \sim 0.15$ and $F_Q(\theta_{23}) \sim 15$, respectively. In contrast, $F_Q(\Delta m_{31}^2)$ increases rapidly with $L/E$ over the range considered, reaching values of order $10^7$. This hierarchy indicates that, within the considered single-parameter framework, the neutrino state carries substantially greater intrinsic quantum sensitivity to $\Delta m_{31}^2$ than to $\theta_{23}$ or $\delta_{\rm CP}$. We further compare the QFI with the classical Fisher information (CFI) obtained from flavor-transition probabilities, showing how much of the available information can be extracted through this specific measurement. 
For a fixed DUNE baseline, matter effects substantially enhance the QFI for $\delta_{\rm CP}$ in the few-GeV region, while producing only minor modifications for $\theta_{23}$ and $\Delta m_{31}^2$. These results show that the information available for estimating the three oscillation parameters differs significantly, and that the flavor-probability measurement does not always extract all the information available in the quantum state. 
}


\begin{document}
\maketitle
\flushbottom


\section{Introduction}
\label{sec:intro}

The discovery of neutrino oscillation has firmly established that neutrinos possess non-zero masses and undergo flavor oscillations as they propagate, implying mixing among flavor and mass eigenstates
~\cite{Super-Kamiokande:1998kpq, SNO:2002tuh, KamLAND:2004mhv, Super-Kamiokande:2016yck}. This phenomenon is described by the Pontecorvo-Maki-Nakagawa-Sakata (PMNS) matrix, which is parameterized by three mixing angles, two independent mass-squared differences, and a CP-violating phase $\delta_{\mathrm{CP}}$~\cite{10.1143/PTP.28.870,Bilenky:1978nj}. Precise determination of these oscillation parameters remains one of the central goals of neutrino physics~\cite{Esteban:2024eli, Rahaman:2022rfp, DiLodovico:2023jgr, T2K:2025wet}. In particular, the leptonic CP-violating phase $\delta_{\mathrm{CP}}$, the atmospheric mixing angle $\theta_{23}$, and the atmospheric mass-squared difference $\Delta m_{31}^{2}$ play a crucial role in determining the oscillation probabilities and their parameter dependence in long-baseline experiments.

The measurement of $\delta_{\mathrm{CP}}$ is of profound importance, as leptonic CP violation may provide key insights into the observed baryon asymmetry of the Universe through mechanisms such as leptogenesis~\cite{Fukugita:1986hr}. At the same time, precise knowledge of $\theta_{23}$ is essential for resolving the octant ambiguity and understanding the structure of lepton mixing. The precise determination of $\Delta m_{31}^2$ is essential for precision neutrino oscillation phenomenology, while its sign is directly related to the neutrino mass ordering. Current and forthcoming long-baseline neutrino experiments, including T2K~\cite{Hu:2024qlx}, NO$\nu$A \cite{Kalitkina:2025hbg}, Hyper-Kamiokande~\cite{Dalmazzone:2025jom}, DUNE~\cite{DUNE:2015lol, Perez-Molina:2026vex}, and ESS$\nu$SB~\cite{Giarnetti:2023pkz, Ghosh:2026cqk}, are expected to significantly improve sensitivity to these parameters. However, challenges such as parameter degeneracies, matter effects, statistical limitations, and systematic uncertainties continue to hinder their precise extraction.

Quantum information science can provide new experimental and theoretical tools for precision measurements in particle physics. In parallel with developments in neutrino phenomenology, concepts and tools from quantum information theory have increasingly been employed to understand and characterize neutrino oscillation dynamics~\cite{Blasone:2007wp, Blasone:2007vw, Blasone:2013zaa, Alok:2014gya, Blasone:2015lya, Banerjee:2015mha, Formaggio:2016cuh, Naikoo:2017fos, Fu:2017hky, Richter-Laskowska:2018ikv, Naikoo:2019eec, Wang:2020vdm, Jha:2020dav, Dixit_2021, Sarkar:2020vob, Shafaq:2020sqo, Bittencourt:2022tcl, Yadav:2022grk, Li:2022mus, Li:2022zic, Jha:2022yik, Ravari:2022yfd, Bittencourt:2023asd, Wang:2023rbf, Chattopadhyay:2023xwr, Dixit:2023fke, Soni:2023njf, Konwar:2024nrd, Konwar:2024nwc, Yadav:2024qav, Konwar:2025ipv}. Treating oscillating neutrinos as evolving quantum states provides a natural framework to investigate their coherence, entanglement properties, and information-theoretic characteristics. Within this approach, quantum Fisher information (QFI) has emerged as a powerful quantity to quantify the ultimate precision limit with which a parameter can be estimated, as dictated by the quantum Cram\'er Rao bound~\cite{Cramer1946}. Unlike classical estimators that depend on specific measurement strategies, QFI provides a measurement-independent bound and thus captures the fundamental  information about a parameter encoded in a quantum state. In contrast, the classical Fisher information (CFI) depends on the chosen measurement and quantifies the information that can be extracted from the outcomes of the chosen measurement. Comparing QFI and CFI therefore provides a way to distinguish the information fundamentally available in the quantum state from that accessible through a specific measurement. It was also used in analyzing collider experiments~\cite{jialiu}.

Applying QFI to neutrino oscillations provides a rigorous framework for quantifying the information encoded in the neutrino quantum state about the fundamental oscillation parameters~\cite{Nogueira:2016qsk, Ignoti:2025rxr}. Evaluating the QFI with respect to $\delta_{\rm CP}$, $\theta_{23}$, and $\Delta m_{31}^2$ allows the intrinsic parameter sensitivity to be characterized as a function of the oscillation kinematics and benchmark oscillation parameters. The corresponding quantum Cram\'er--Rao bound establishes the ultimate precision permitted by the quantum state for single-parameter estimation. The comparison with the CFI further allows us to assess how much of this quantum information is accessible through the considered flavor-transition measurements. In addition, matter effects modify the neutrino evolution and can consequently alter the parameter sensitivity in a parameter-dependent manner.

In this work, we investigate the QFI for the three parameters $\delta_{\rm CP}$, $\theta_{23}$, and $\Delta m_{31}^2$, treating one parameter at a time within the three-flavor neutrino oscillation framework. We first analyze the evolution of the neutrino state in vacuum and study the $L/E$ dependence of the QFI using the NuFit-6.0 benchmark parameter sets. We then calculate the CFI associated with the flavor-transition probabilities $P(\nu_\mu \to \nu_e)$, $P(\nu_\mu \to \nu_\mu)$, and $P(\nu_\mu \to \nu_\tau)$, and compare it with the corresponding QFI to quantify the information accessible through the considered flavor measurement. We further investigate the impact of matter effects for a DUNE-like baseline and examine the dependence of the QFI on the neutrino mass ordering. This combined analysis distinguishes the fundamental quantum information available in the evolving neutrino state from the information accessible through a specific flavor measurement, while illustrating the parameter-dependent influence of matter effects on quantum sensitivity. 

The paper is organized as follows. In Sec.~\ref{sec:neutrino}, we briefly review the three-flavor neutrino oscillation framework and the evolution of the neutrino quantum state in vacuum and in the presence of matter. In Sec.~\ref{sec:fi}, we introduce the quantum and classical Fisher information, and the corresponding Cram\'er--Rao bounds, emphasizing the distinction between the fundamental quantum information available in the state and the information accessible through a specific flavor measurement. In Sec.~\ref{sec:result}, we present our numerical results for the QFI of $\delta_{\rm CP}$, $\theta_{23}$, and $\Delta m_{31}^2$ as functions of $L/E$, using the NuFit-6.0 benchmark parameter sets. We then compare the QFI with the CFI obtained from flavor-transition probabilities and investigate the impact of matter effects for a DUNE-like baseline. The dependence of the QFI on the neutrino mass ordering is also examined. Finally, in Sec.~\ref{sec:con}, we summarize our main findings and discuss possible future extensions to multiparameter estimation using the quantum Fisher information matrix, including correlations among the oscillation parameters.

\section{Neutrino Oscillation}
\label{sec:neutrino}
Neutrino oscillation is a quantum mechanical phenomenon arising from the mismatch between neutrino flavor eigenstates and mass eigenstates~\cite{Giunti:2007ry, Ohlsson:1999xb}. Neutrinos are produced and detected in flavor eigenstates $|\nu_\alpha\rangle$, with $\alpha=e,\mu,\tau$, which are linear superpositions of the mass eigenstates $|\nu_i\rangle$, with $i=1,2,3$ and definite masses $m_i$. The mixing between flavor and mass eigenstates is described by the unitary Pontecorvo--Maki--Nakagawa--Sakata (PMNS) matrix $U$, such that
\begin{align}
|\nu_\alpha\rangle
=
\sum_{i=1}^{3} U_{\alpha i} |\nu_i\rangle .
\end{align}

As neutrinos propagate, the mass eigenstates acquire different quantum phases because of their distinct masses. For ultra-relativistic neutrinos with energy $E$, the evolution of the $i$-th mass eigenstate can be written as
\begin{align}
|\nu_i(t)\rangle
&=
e^{-i(E_i t-p_iL)}|\nu_i\rangle
\nonumber\\
&\simeq
\exp\left(-i\frac{m_i^2L}{2E}\right)|\nu_i\rangle ,
\end{align}
where $L\simeq t$ in natural units. The relative phases accumulated by the mass eigenstates therefore lead to a distance-dependent change in the flavor composition of the neutrino state.
\subsection{ Vacuum evolution}
Neutrino propagation in vacuum can be described by a Schr\"odinger-like evolution equation,
\begin{align}
i\frac{d}{dL}|\nu(L)\rangle
=
H_{\rm vac}|\nu(L)\rangle ,
\end{align}
where, in the flavor basis, the vacuum Hamiltonian is
\begin{align}
H_{\rm vac}
=
\frac{1}{2E}
U
\begin{pmatrix}
m_1^2 & 0 & 0\\
0 & m_2^2 & 0\\
0 & 0 & m_3^2
\end{pmatrix}
U^\dagger .
\label{eq:Hvac}
\end{align}

Since a term proportional to the identity contributes only a common phase to the neutrino state, which has no effect on flavor-transition probabilities or the QFI, the Hamiltonian can equivalently be expressed in terms of the mass-squared differences as
\begin{align}
H_{\rm vac}
=
\frac{1}{2E}
U
\begin{pmatrix}
0 & 0 & 0\\
0 & \Delta m_{21}^2 & 0\\
0 & 0 & \Delta m_{31}^2
\end{pmatrix}
U^\dagger,
\label{eq:Hvac_delta}
\end{align}
where
\begin{align}
\Delta m_{ij}^{2}=m_i^2-m_j^2 .
\end{align}

For constant vacuum propagation, the corresponding evolution operator is
\begin{align}
S_{\rm vac}(L)
=
e^{-iH_{\rm vac}L},
\end{align}
and an initially produced flavor state evolves according to
\begin{align}
|\nu_\alpha(L)\rangle
=
S_{\rm vac}(L)|\nu_\alpha(0)\rangle .
\label{eq:state_vac}
\end{align}

The probability of detecting a neutrino produced in initial flavor $\nu_\alpha$ as final flavor $\nu_\beta$ after propagation over a distance $L$ is given by~\cite{Giunti:2007ry,Yasuda:2007jp}
\begin{align}
P_{\alpha\rightarrow\beta}
&=
\delta_{\alpha\beta}
-4\sum_{i>j}
\mathrm{Re}\left(
U_{\alpha i}U_{\beta i}^{\ast}
U_{\alpha j}^{\ast}U_{\beta j}
\right)
\sin^{2}\left(
\frac{\Delta m_{ij}^{2}L}{4E}
\right)
\nonumber\\
&\quad+
2\sum_{i>j}
\mathrm{Im}\left(
U_{\alpha i}U_{\beta i}^{\ast}
U_{\alpha j}^{\ast}U_{\beta j}
\right)
\sin\left(
\frac{\Delta m_{ij}^{2}L}{2E}
\right).
\label{eq:prob_vac}
\end{align}

The first two terms describe the CP-even oscillation contribution, while the term proportional to the imaginary part contains the CP-odd contribution and hence the dependence on the leptonic CP-violating phase $\delta_{\rm CP}$. The oscillation probabilities are therefore sensitive to the mixing angles $\theta_{ij}$, the mass-squared differences $\Delta m_{21}^2$ and $\Delta m_{31}^2$, and the CP-violating phase.

\subsection{Matter effects}

For long-baseline neutrino propagation, coherent forward scattering with electrons in matter introduces an additional effective potential for electron neutrinos. For approximately constant matter density, the neutrino evolution is governed by the effective Hamiltonian \cite{Ohlsson:1999xb,Wolfenstein:1977ue}
\begin{align}
H_{\rm mat}
=
H_{\rm vac}
+
\begin{pmatrix}
V_{\rm CC} & 0 & 0\\
0 & 0 & 0\\
0 & 0 & 0
\end{pmatrix},
\label{eq:Hmatter}
\end{align}
where
\begin{align}
V_{\rm CC}
=
\sqrt{2}\,G_F N_e ,
\label{eq:Vcc}
\end{align}
with $G_F$ denoting the Fermi constant and $N_e$ the electron number density. The additional potential modifies the flavor evolution and consequently the oscillation probabilities and their dependence on the underlying oscillation parameters.

For constant matter density, the corresponding evolution operator is
\begin{align}
S_{\rm mat}(L)
=
e^{-iH_{\rm mat}L},
\end{align}
and the evolved state is
\begin{align}
|\nu_\alpha(L)\rangle_{\rm mat}
=
S_{\rm mat}(L)|\nu_\alpha(0)\rangle .
\label{eq:state_matter}
\end{align}

The transition probability in matter is consequently given by
\begin{align}
P_{\alpha\rightarrow\beta}^{\rm mat}
=
\left|
\langle\nu_\beta|
S_{\rm mat}(L)
|\nu_\alpha(0)\rangle
\right|^2 .
\label{eq:prob_matter}
\end{align}

In this work, this matter-modified evolution is used to examine how propagation through a DUNE-like matter environment changes the Fisher information relative to vacuum propagation.

The vacuum and matter evolution described above provide the parameter-dependent quantum states used in the quantum-information analysis. In the following section, we introduce the quantum and classical Fisher information  to quantify, respectively, the information available in the evolving quantum state and the information accessible through the considered flavor-transition measurements.

\section{Fisher Information}
\label{sec:fi}

Quantum Fisher Information provides a quantitative measure of the
information about an unknown parameter encoded in a quantum state. It
therefore sets the ultimate precision bound for parameter estimation,
independent of any particular measurement scheme. For a family of density
matrices $\rho(\lambda)$ depending on a parameter $\lambda$, the quantum
Cram\'er--Rao bound gives the lower bound on the variance of any unbiased
estimator $\hat{\lambda}$ as~\cite{Cramer1946}
\begin{equation}
(\Delta\lambda)^2
\geq
\frac{1}{M F_Q(\lambda)},
\label{eq:cramer}
\end{equation}
where $M$ denotes the number of independent measurements and
$F_Q(\lambda)$ is the QFI associated with the parameter $\lambda$.
Thus, a larger QFI corresponds to a smaller quantum-limited uncertainty
in the estimation of $\lambda$. Importantly, this bound represents an
intrinsic property of the parameter-dependent quantum state and does not
include experimental effects such as finite event statistics, detector
efficiencies, backgrounds, or systematic uncertainties.

The QFI can be defined through the symmetric logarithmic derivative
(SLD) operator $L_\lambda$, which satisfies~\cite{Braunstein:1994zz,
Paris:2008zgg, Jacobs_2014, Toth:2014msl}
\begin{equation}
\frac{\partial\rho(\lambda)}{\partial\lambda}
=
\frac{1}{2}
\left[
\rho(\lambda)L_\lambda
+
L_\lambda\rho(\lambda)
\right].
\end{equation}
The QFI is then given by
\begin{equation}
F_Q(\lambda)
=
\mathrm{Tr}\left[\rho(\lambda)L_\lambda^2\right].
\label{eq:qfi_sld}
\end{equation}

In the present analysis, the neutrino is treated as an evolving pure
quantum state,
\begin{equation}
\rho(\lambda)
=
|\psi(\lambda)\rangle
\langle\psi(\lambda)|,
\end{equation}
for which the QFI takes the simpler form~\cite{helstrom_quantum_1976,
Braunstein:1994zz, Paris:2008zgg, RevModPhys.90.035005}
\begin{equation}
F_Q(\lambda)
=
4\left[
\langle\partial_\lambda\psi|
\partial_\lambda\psi\rangle
-
\left|
\langle\psi|\partial_\lambda\psi\rangle
\right|^2
\right].
\label{eq:qfi_pure}
\end{equation}
This expression quantifies the sensitivity of the quantum state to
variations in the parameter $\lambda$. A larger value of $F_Q$ indicates
that quantum states corresponding to nearby values of $\lambda$ are more
distinguishable and, consequently, that a smaller uncertainty is
permitted by the quantum Cram\'er--Rao bound. In this work, we consider
single-parameter estimation with
\begin{equation}
\lambda\in
\left\{
\delta_{\rm CP},\,
\theta_{23},\,
\Delta m_{31}^{2}
\right\}.
\end{equation}

While the QFI characterizes the maximum information available in the
quantum state, an actual measurement produces a probability distribution
over a set of measurement outcomes. For a measurement described by
positive-operator-valued measures (POVMs) $\{\Pi_x\}$, the probability of
obtaining outcome $x$ is
\begin{equation}
p(x|\lambda)
=
\mathrm{Tr}\left[\rho(\lambda)\Pi_x\right].
\end{equation}
The corresponding classical Fisher information (CFI) is
\begin{equation}
F_C(\lambda)
=
\sum_x
\frac{1}{p(x|\lambda)}
\left(
\frac{\partial p(x|\lambda)}
{\partial\lambda}
\right)^2 .
\label{eq:cfi_general}
\end{equation}
Unlike the QFI, the CFI depends explicitly on the chosen measurement,
since different measurements can produce different probability
distributions and thus different amounts of accessible information.

For neutrino oscillations, we consider flavor detection as the
measurement scheme. For an initial flavor $\nu_\alpha$, the possible
measurement outcomes are the detected flavors $\nu_\beta$, with the
corresponding probabilities given by the flavor-transition probabilities
$P_{\alpha\rightarrow\beta}$. The CFI therefore becomes
\begin{equation}
F_C(\lambda)
=
\sum_{\beta=e,\mu,\tau}
\frac{1}{P_{\alpha\rightarrow\beta}}
\left(
\frac{\partial P_{\alpha\rightarrow\beta}}
{\partial\lambda}
\right)^2 .
\label{eq:cfi}
\end{equation}
For the $\nu_\mu\rightarrow\nu_e$ oscillation channel considered in this
work, the three possible flavor outcomes are $\nu_e$, $\nu_\mu$, and
$\nu_\tau$, and hence
\begin{equation}
F_C(\lambda)
=
\frac{1}{P_{\mu e}}
\left(
\frac{\partial P_{\mu e}}{\partial\lambda}
\right)^2
+
\frac{1}{P_{\mu\mu}}
\left(
\frac{\partial P_{\mu\mu}}{\partial\lambda}
\right)^2
+
\frac{1}{P_{\mu\tau}}
\left(
\frac{\partial P_{\mu\tau}}{\partial\lambda}
\right)^2 .
\label{eq:cfi_numu}
\end{equation}
The probabilities satisfy
\begin{equation}
P_{\mu e}+P_{\mu\mu}+P_{\mu\tau}=1,
\end{equation}
so that these three flavor outcomes form the complete set of outcomes
for the considered idealized three-flavor measurement.

For any quantum state and any specified measurement, the classical
Fisher information is bounded from above by the QFI,
\begin{equation}
F_C(\lambda)\leq F_Q(\lambda).
\label{eq:qfi_cfi}
\end{equation}
This inequality provides the central connection between the two
quantities. The QFI characterizes the maximum information about
$\lambda$ encoded in the quantum state, whereas the CFI quantifies the
information extracted by the specific flavor measurement considered
here. Consequently, a difference between $F_Q$ and $F_C$ indicates that
the chosen measurement does not access all of the information available
in the quantum state. If an appropriate measurement saturates the bound,
$F_C=F_Q$, the corresponding measurement reaches the QFI limit. The comparison of QFI and CFI therefore allows us to
quantify the measurement dependence of the accessible information in
neutrino oscillation parameter estimation.

\begin{table}[!htb]
\centering
\caption{Best-fit values and $1\sigma$ ranges of the three-flavor neutrino oscillation parameters from NuFit-6.0 global analysis for the two datasets considered in this work \cite{Esteban:2024eli}.}
\label{tab:nufit_bestfit_1sigma}
\renewcommand\arraystretch{1.5}
\begin{center}
	\hspace{0.05cm}
	   \renewcommand\arraystretch{1.5}
		  \scalebox{0.75}
		  {
\begin{tabular}{c|cc|cc}
\hline\hline
 & \multicolumn{2}{c|}{\textbf{Normal Ordering (NO)}} 
 & \multicolumn{2}{c}{\textbf{Inverted Ordering (IO)}} \\

\textbf{Parameter} 
 & Best fit $\pm 1\sigma$  & Best fit $\pm 1\sigma$  & Best fit $\pm 1\sigma$   & Best fit $\pm 1\sigma$ )\\
  &(IC19 without SK-atm) &  (IC24 with SK-atm) &  (IC19 without SK-atm) &  (IC24 with SK-atm) \\
\hline
$\theta_{12}\,[^\circ]$ & $33.68^{+0.73}_{-0.70}$ & $33.68^{+0.73}_{-0.70}$ & $33.68^{+0.73}_{-0.70}$ & $33.68^{+0.73}_{-0.70}$ \\

$\theta_{23}\,[^\circ]$ & $48.5^{+0.7}_{-0.9}$  & $43.3^{+1.0}_{-0.8}$ & $48.6^{+0.7}_{-0.9}$ 
& $47.9^{+0.7}_{-0.9}$ \\

$\theta_{13}\,[^\circ]$ & $8.52^{+0.11}_{-0.11}$ & $8.56^{+0.11}_{-0.11}$ & $8.58^{+0.11}_{-0.11}$ & $8.59^{+0.11}_{-0.11}$ \\

$\delta_{\rm CP}\,[^\circ]$ & $177^{+19}_{-20}$ & $212^{+26}_{-41}$  & $285^{+25}_{-28}$ 
& $274^{+25}_{-23}$ \\

$\Delta m^2_{21}\,[10^{-5}\,\mathrm{eV}^2]$ & $7.49^{+0.19}_{-0.19}$ & $7.49^{+0.19}_{-0.19}$ 
 & $7.49^{+0.19}_{-0.19}$ & $7.49^{+0.19}_{-0.19}$ \\

$\Delta m^2_{31}\,[10^{-3}\,\mathrm{eV}^2]$ & $+2.534^{+0.025}_{-0.023}$ & $+2.513^{+0.021}_{-0.019}$ & $-2.510^{+0.024}_{-0.025}$ & $-2.484^{+0.020}_{-0.020}$ 
\\
\hline\hline
\end{tabular}
}
\end{center}
\end{table}

\section{Results and discussion}
\label{sec:result}

We now present the results for the quantum and classical Fisher
information associated with the individual estimation of the oscillation
parameters $\delta_{\mathrm{CP}}$, $\theta_{23}$, and $\Delta m_{31}^{2}$.
For each parameter, we vary the parameter of interest around its benchmark
value while keeping the remaining oscillation parameters fixed. This
allows us to isolate the dependence of the evolving neutrino quantum state
on a single parameter at a time. The QFI is evaluated as a function of the
baseline-to-energy ratio $L/E$, providing a direct characterization of the
parameter-dependent information encoded in the neutrino state.

We focus on the $\nu_\mu\rightarrow\nu_e$ appearance channel and, for the
classical Fisher information, consider the complete set of flavor outcomes
$P(\nu_\mu\rightarrow\nu_e)$, $P(\nu_\mu\rightarrow\nu_\mu)$, and
$P(\nu_\mu\rightarrow\nu_\tau)$. The appearance probability
$P(\nu_\mu\rightarrow\nu_e)$ has a particularly important role in
long-baseline oscillation experiments because of its dependence on
$\delta_{\mathrm{CP}}$ and $\theta_{23}$ and its sensitivity to matter
effects. Comparing the CFI obtained from the flavor-transition
probabilities with the corresponding QFI allows us to distinguish the
information available in the neutrino quantum state from the information
accessible through the considered flavor measurement.

Unless stated otherwise, the oscillation parameters are fixed to the
benchmark values listed in Table~\ref{tab:nufit_bestfit_1sigma}, consistent
with the NuFit-6.0 global fits. We consider both the IC24 benchmark,
including Super-Kamiokande atmospheric data, and the IC19 benchmark,
without Super-Kamiokande atmospheric data, where applicable. This allows
us to examine the robustness of the QFI results with respect to the choice
of global-fit dataset.

To connect the $L/E$ dependence of the Fisher information with the
kinematic regimes probed by long-baseline experiments, we indicate the
representative $L/E$ values of MINOS, T2K, NO$\nu$A, DUNE, and ESS$\nu$SB
in the corresponding figures. Here, $L$ denotes the experimental baseline
and $E$ is taken to be the neutrino energy at which the flux is maximal for
the corresponding facility; the values used in the analysis are summarized
in Appendix~\ref{app:exp}. These benchmark points provide a reference for
interpreting the quantum and classical information available in the
$L/E$ regions relevant to existing and proposed experiments.

\begin{figure}[!htb]
\centering
\includegraphics[width=0.75\textwidth]{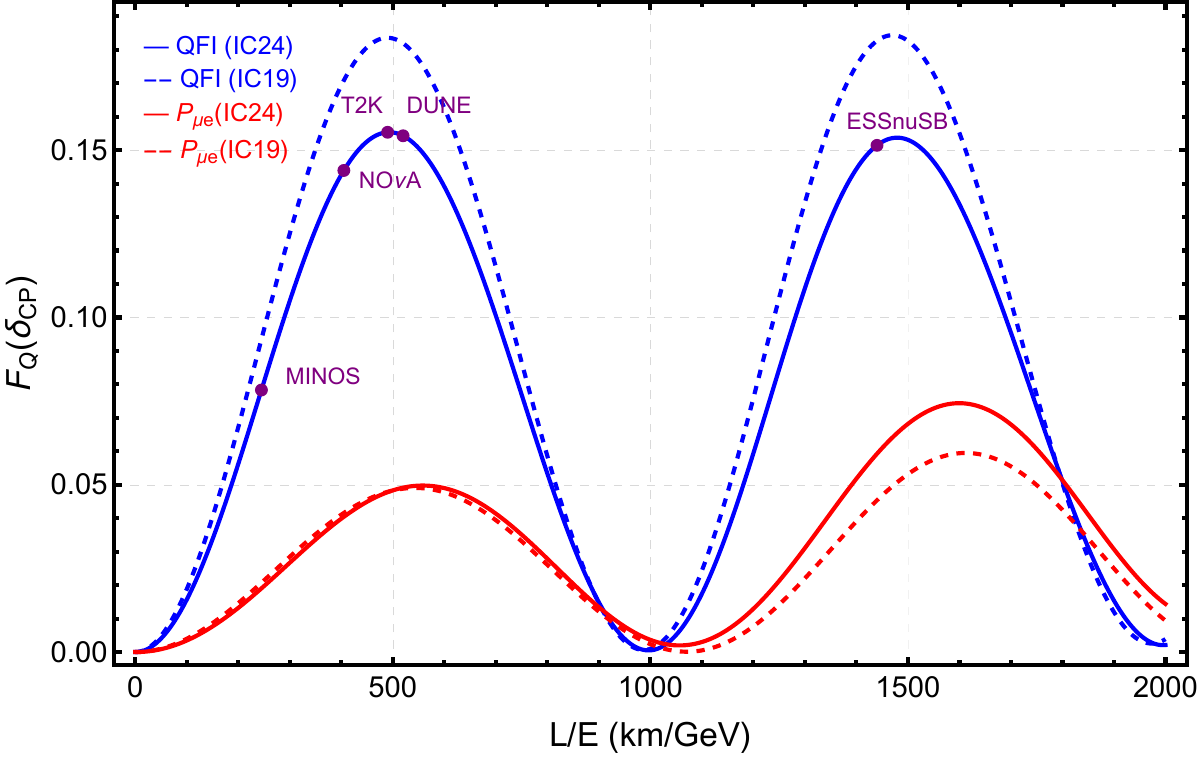}
\caption{
QFI ${F}_Q(\delta_{\rm CP})$ and the oscillation probability $P(\nu_\mu \to \nu_e)$ as functions of $L/E$. The blue curves show the QFI, while the red curves represent the transition probability. Solid lines correspond to the NuFit-6.0 IC24 dataset including Super-Kamiokande (S-K) atmospheric data, whereas dashed lines correspond to the IC19 dataset without S-K data.}
\label{fig:dcp}
\end{figure}

Figure~\ref{fig:dcp} shows the QFI for $\delta_{\mathrm{CP}}$ evaluated
at two benchmark sets of oscillation parameters: NuFit IC24 including
Super-Kamiokande (S-K) atmospheric data (solid lines) and IC19 without
atmospheric data (dashed lines), plotted as a function of $L/E$. The QFI
exhibits two prominent maxima at $L/E \sim 500~\mathrm{km/GeV}$ and
$L/E \sim 1500~\mathrm{km/GeV}$, which closely coincide with the first and
second maxima of the transition probability
$P(\nu_\mu\rightarrow\nu_e)$. This correspondence indicates that the
regions of maximal $\nu_\mu\rightarrow\nu_e$ transition probability also
provide enhanced intrinsic sensitivity of the evolving quantum state to
$\delta_{\mathrm{CP}}$. The QFI reaches values of approximately
$F_Q(\delta_{\mathrm{CP}})\sim 0.15$ in these regions. The representative
$L/E$ values of T2K, NO$\nu$A, DUNE, and ESS$\nu$SB are also indicated,
showing that the first and second sensitivity maxima lie close to the
kinematic regimes probed by current and proposed long-baseline
experiments. Only minor differences are observed between the IC24 and
IC19 results, indicating that the QFI sensitivity to $\delta_{\mathrm{CP}}$
is relatively robust against the inclusion of S-K atmospheric data in the
global-fit benchmark. The overall behavior is also consistent with the
QFI dependence on $\delta_{\mathrm{CP}}$ reported in
Ref.~\cite{Ignoti:2025rxr}.

\begin{figure}[!htb]
\centering
\includegraphics[width=0.75\textwidth]{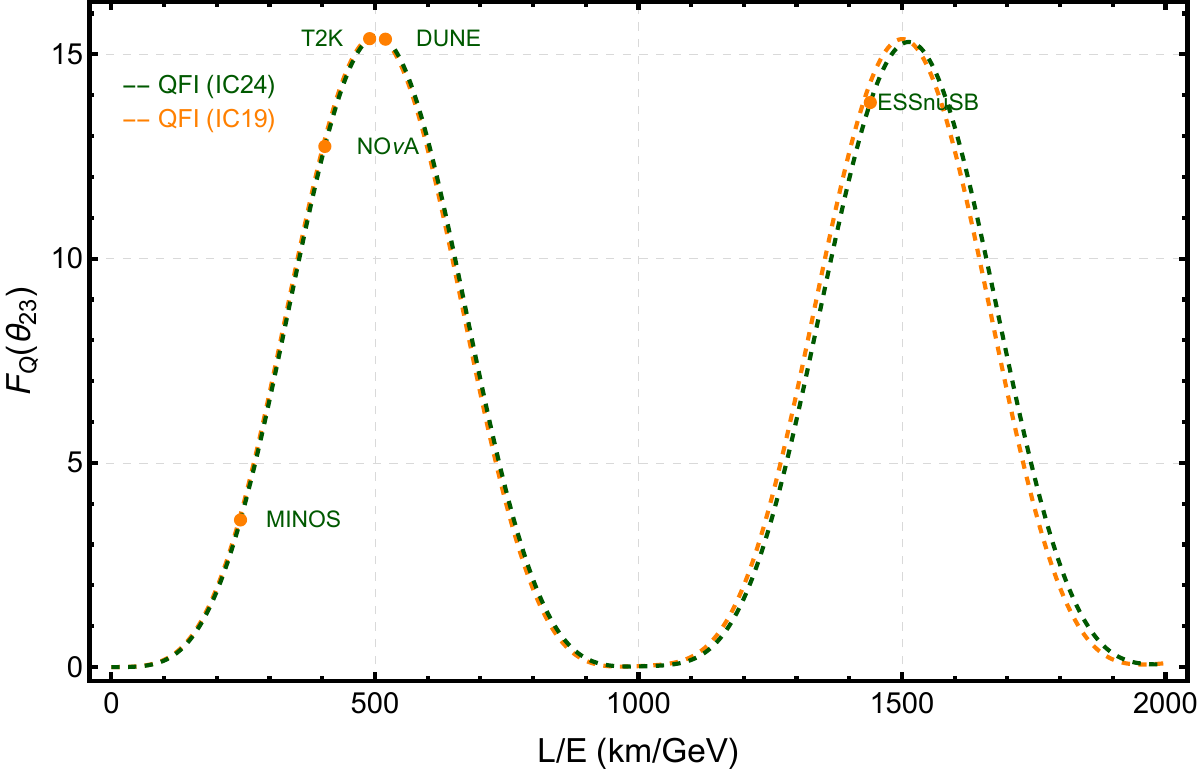}
\caption{QFI for $\Theta_{\rm 23}$ as functions of $L/E$. The two curves show QFI where dashed green line correspond to the NuFit-6.0 IC24 dataset including S-K atmospheric data, whereas dashed orange line correspond to the IC19 dataset without S-K data.}
\label{fig:theta23}
\end{figure}

The atmospheric mixing angle $\theta_{23}$ exhibits a qualitatively
similar QFI behavior to $\delta_{\mathrm{CP}}$, with a bimodal structure
featuring maxima at $L/E \sim 500~\mathrm{km/GeV}$ and
$L/E \sim 1500~\mathrm{km/GeV}$, as shown in Fig.~\ref{fig:theta23}.
However, the quantum sensitivity to $\theta_{23}$ is substantially larger,
with the QFI reaching $F_Q(\theta_{23})\sim 15$, approximately two orders
of magnitude above the corresponding maximum
$F_Q(\delta_{\mathrm{CP}})\sim0.15$. Through the quantum Cram\'er--Rao
bound in Eq.~\eqref{eq:cramer}, this larger QFI corresponds to a tighter
quantum-limited precision bound on $\theta_{23}$. The two benchmark
datasets, IC24 (solid) and IC19 (dashed), yield nearly indistinguishable
QFI profiles, indicating that the intrinsic quantum sensitivity to
$\theta_{23}$ is robust against the inclusion of Super-Kamiokande
atmospheric data in the global-fit parameter set. The representative
$L/E$ values of T2K, NO$\nu$A, and DUNE lie close to the first QFI maximum,
while ESS$\nu$SB lies near the second maximum. Thus, the two oscillation
regimes provide comparable intrinsic quantum sensitivity to
$\theta_{23}$, illustrating the complementary $L/E$ regions probed by
different long-baseline configurations. As in the case of
$\delta_{\mathrm{CP}}$, these QFI maxima characterize the information
encoded in the neutrino quantum state and should not be interpreted as
direct measures of experimental sensitivity, which additionally depends
on event statistics, detector response, and systematic uncertainties.

\begin{figure}[!htb]
\centering
\includegraphics[width=0.75\textwidth]{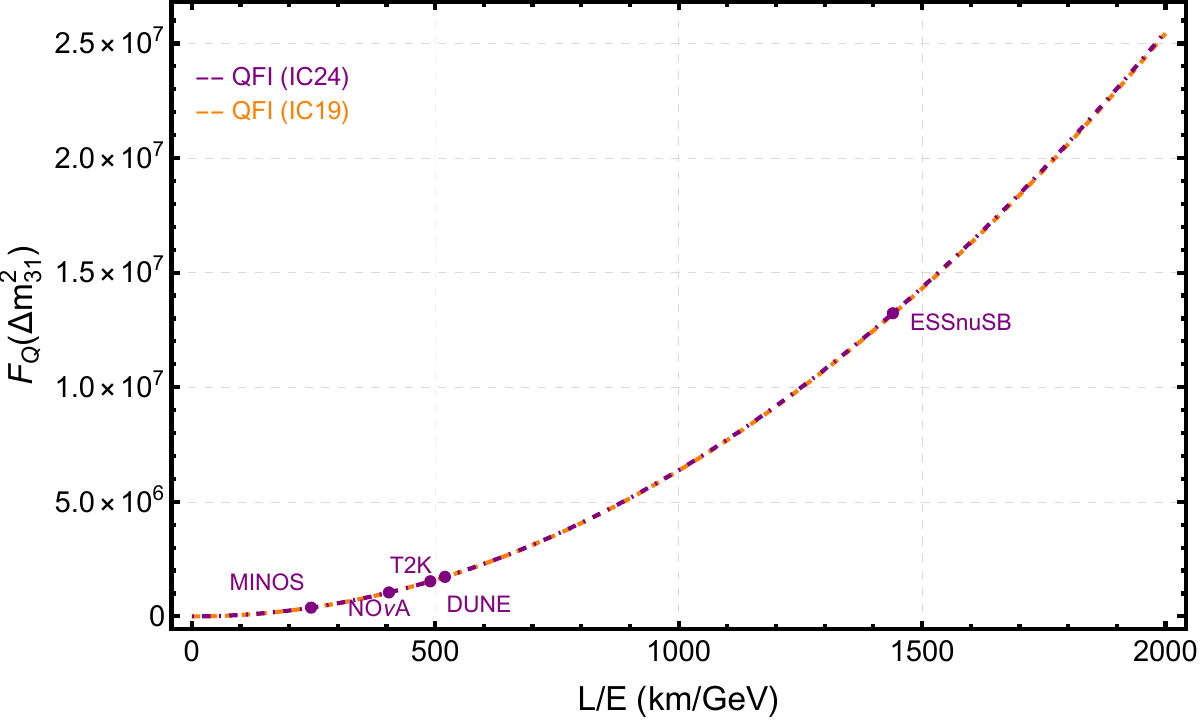}
\caption{QFI for $\Delta m_{\rm 31}^{2}$ as function of $L/E$. The two curves show QFI, where dashed purple line correspond to the NuFit-6.0 IC24 dataset including S-K atmospheric data, whereas dashed orange line correspond to the IC19 dataset without S-K data.}
\label{fig:dm31}
\end{figure}

{\color{black}The mass-squared difference $\Delta m_{31}^2$ exhibits a markedly different QFI behavior compared with the angular parameters, as shown in Fig.~\ref{fig:dm31}. In contrast to the bimodal structures observed for $\delta_{\rm CP}$ and $\theta_{23}$, the QFI for $\Delta m_{31}^2$ increases rapidly with $L/E$ over the range considered, reaching values of order $2.5\times10^7$ near $L/E = 2000~\mathrm{km/GeV}$. This substantially larger QFI reflects the distinct role of $\Delta m_{31}^2$ in the oscillation dynamics. While the angular parameters primarily modify the amplitudes and interference structure of the oscillation probabilities, $\Delta m_{31}^2$ directly controls the oscillation phase, making the quantum state increasingly sensitive to variations in $\Delta m_{31}^2$ as $L/E$ increases. The corresponding quantum bound therefore indicates a much higher intrinsic information content for $\Delta m_{31}^2$ than for the angular parameters within the considered framework. The benchmark $L/E$ values for MINOS, T2K, NO$\nu$A, DUNE, and ESS$\nu$SB lie along the same rapidly increasing QFI profile, with the ESS$\nu$SB benchmark located in a region of substantially larger intrinsic quantum sensitivity. However, this larger QFI should not by itself be interpreted as superior experimental precision, since the experimentally attainable sensitivity also depends on event statistics, detector response, and systematic uncertainties. Finally, the near-complete overlap between the IC24 and IC19 curves demonstrates that the QFI sensitivity to $\Delta m_{31}^2$ is robust against the inclusion of Super-Kamiokande atmospheric data in the global-fit parameter set.}

\begin{figure}[!htb]
\centering
\includegraphics[width=0.75\textwidth]{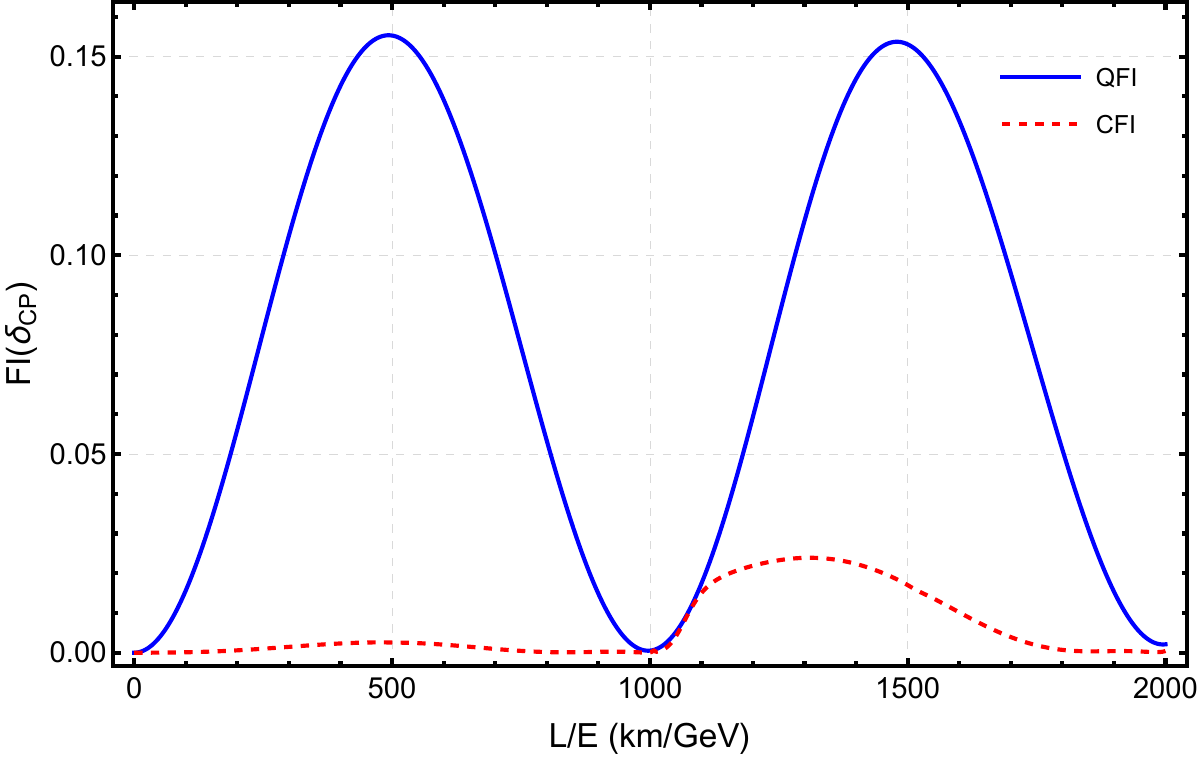}
\caption{Comparison of the quantum Fisher information $F_Q(\delta_{\rm CP})$ and classical Fisher information $F_C(\delta_{\rm CP})$ as functions of $L/E$ for the NuFit-6.0 IC24 dataset, including Super-Kamiokande atmospheric data. The blue solid and red dashed curves represent the QFI and CFI, respectively. The difference between the two quantifies the information not captured by the considered flavor-probability measurement. }
\label{fig:4}
\end{figure}

{\color{black} To complement the quantum-information analysis, we calculate the classical Fisher information associated with the flavor-transition probabilities
$P(\nu_\mu \to \nu_e)$, $P(\nu_\mu \to \nu_\mu)$, and $P(\nu_\mu \to \nu_\tau)$, as shown in Fig.~\ref{fig:4}.
The CFI exhibits an oscillatory structure with enhanced sensitivity in the $L/E$ regions where the QFI reaches its maxima. However, the maxima of the CFI do not coincide exactly with those of the QFI, with a small displacement particularly evident around the second oscillation maximum. This difference reflects the measurement dependence of the CFI, which is determined by the response of the selected flavor-probability outcomes to variations in $\delta_{\rm CP}$. Moreover, $F_C$ remains substantially below $F_Q$ over most of the $L/E$ range, indicating that the considered flavor measurement does not fully extract the information encoded in the quantum state. While $F_Q$ represents the ultimate quantum limit on the estimation of $\delta_{\rm CP}$, $F_C$ quantifies the information accessible through the specific flavor-probability measurement considered here. The difference between the two therefore indicates potential for improved parameter sensitivity through an optimized measurement strategy.}

\begin{figure}[!htb]
\centering
\includegraphics[width=0.75\textwidth]{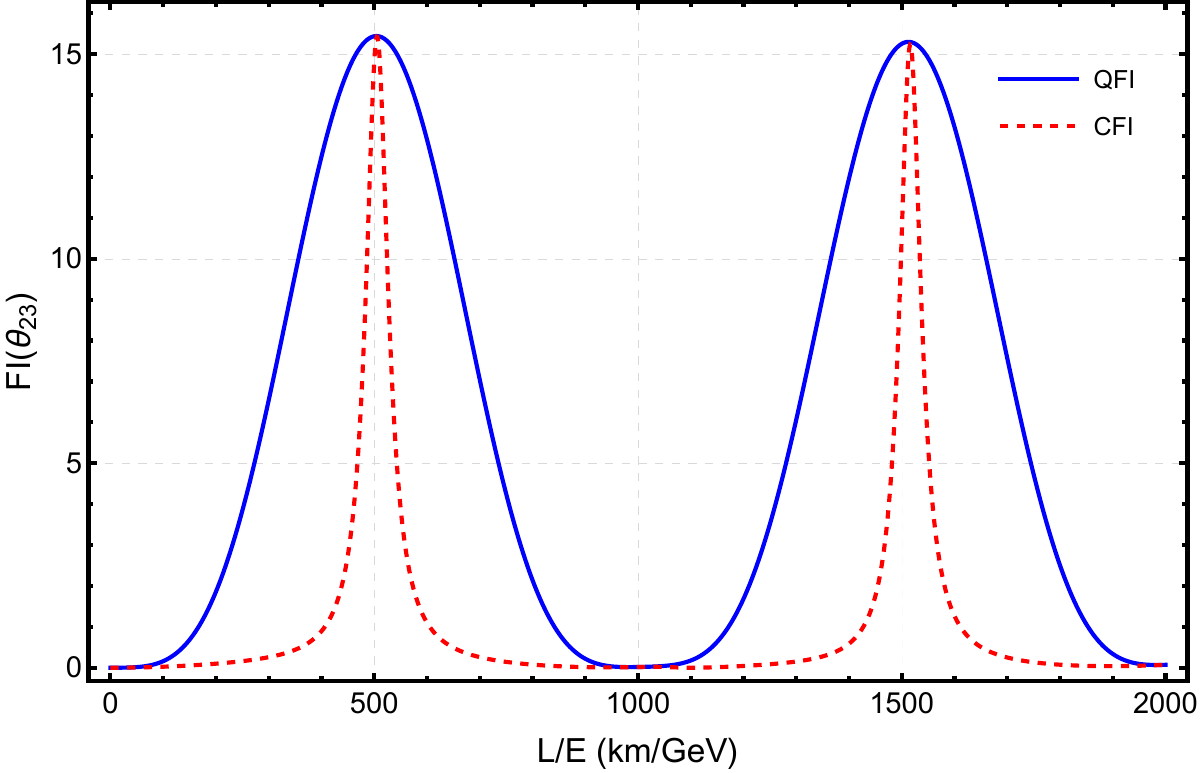}
\caption{Quantum Fisher information $F_Q(\theta_{23})$ and classical Fisher information $F_C(\theta_{23})$ as functions of $L/E$ in vacuum for the NuFit-6.0 IC24 dataset including Super-Kamiokande atmospheric data. The blue solid and red dotted curves represent the QFI and CFI, respectively. The close agreement near the sensitivity maxima indicates that the considered flavor-probability measurement captures a substantial fraction of the available quantum information on $\theta_{23}$. }
\label{fig:5}
\end{figure}

\begin{figure}[!htb]
\centering
\includegraphics[width=0.75\textwidth]{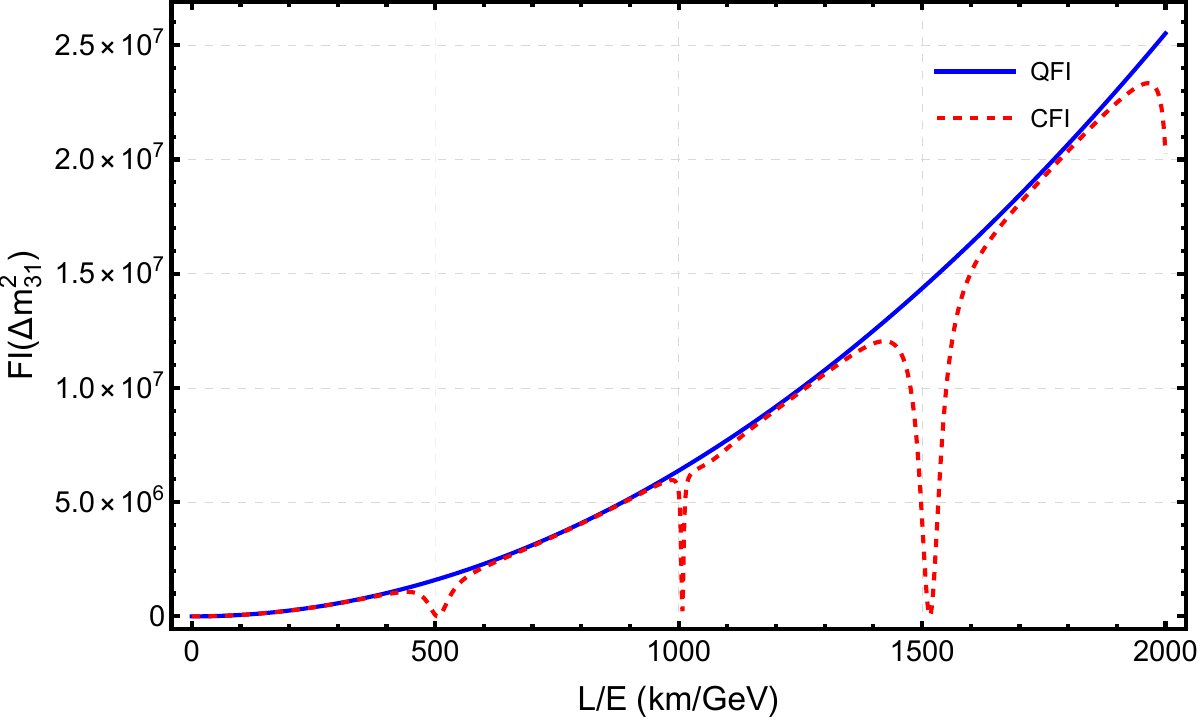}
\caption{Quantum Fisher information $F_Q(\Delta m_{31}^2)$ and classical Fisher information $F_C(\Delta m_{31}^2)$ as functions of $L/E$ in vacuum for the NuFit-6.0 IC24 dataset including Super-Kamiokande atmospheric data. The blue solid and red dotted curves represent the QFI and CFI, respectively. The close correspondence between $F_Q$ and $F_C$ over a broad $L/E$ range indicates that the considered flavor-probability measurement captures most of the available quantum information on $\Delta m_{31}^2$, with localized deviations around the oscillation-sensitive regions.  }
\label{fig:6}
\end{figure}

{\color{black}To further examine the measurement dependence of the available information, we compare the classical and quantum Fisher information for $\theta_{23}$, as shown in Fig.~\ref{fig:5}. The QFI exhibits the characteristic bimodal structure, with maxima around $L/E \sim 500$ and $1500~\mathrm{km/GeV}$. Interestingly, the CFI reaches values very close to the corresponding QFI maxima in these regions, indicating that the considered flavor-probability measurement can extract a large fraction of the available quantum information on $\theta_{23}$ near the optimal sensitivity regions. Away from these regions, however, the CFI decreases more rapidly than the QFI, leading to a pronounced separation between the two. This behavior contrasts with the $\delta_{\rm CP}$ case, where the CFI remains substantially below the QFI over a broader $L/E$ range. The result suggests that the standard flavor-probability measurement is particularly effective for estimating $\theta_{23}$ near the oscillation-sensitivity maxima, whereas further optimization of the measurement strategy may be more relevant away from these regions.}

{\color{black} We further compare the QFI and CFI for $\Delta m_{31}^2$, as shown in Fig.~\ref{fig:6}. The QFI increases smoothly with $L/E$ over the considered range, reflecting the strong sensitivity of the oscillation phase to the atmospheric mass-squared splitting. The CFI closely follows the QFI over a substantial portion of the $L/E$ range, indicating that the flavor-probability measurement captures most of the available information on $\Delta m_{31}^2$. However, pronounced reductions in the CFI are observed around $L/E \sim 500$, $1000$ and $1500~\mathrm{km/GeV}$, where the CFI departs significantly from the QFI. Thus, despite the overall close correspondence between the two quantities, the flavor measurement does not uniformly saturate the quantum information bound. These localized features indicate that the information extracted by the specific flavor measurement depends sensitively on the $L/E$-dependent flavor-transition probabilities. The relatively small QFI--CFI difference over broad regions nevertheless indicates that the standard flavor-probability measurement is highly effective for extracting information on $\Delta m_{31}^2$.}

\begin{figure}[!htb]
\centering
\includegraphics[width=0.75\textwidth]{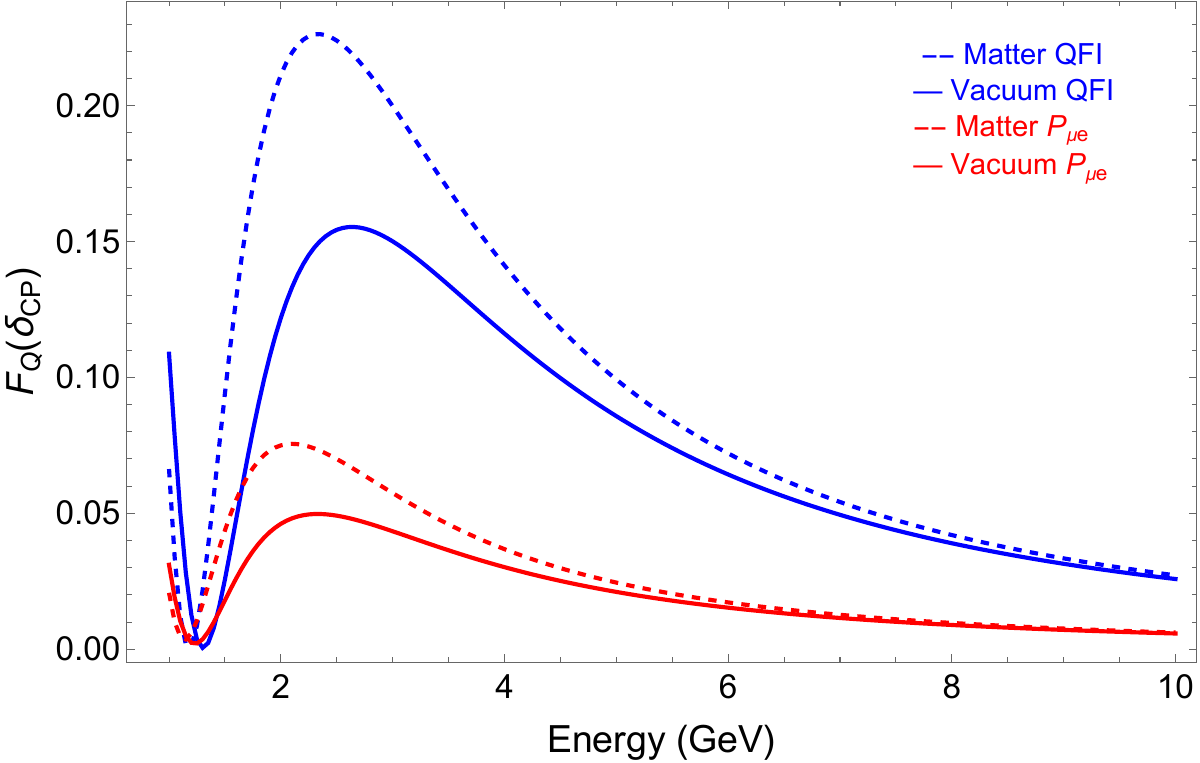}
\caption{QFI $F_Q(\delta_{\rm CP})$ and the oscillation probability $P(\nu_\mu \to \nu_e)$ as functions of neutrino energy $E$ for a fixed baseline of $L = 1300~\mathrm{km}$. The blue curves represent the QFI, while the red curves show the transition probability. Solid curves correspond to vacuum propagation, whereas dashed curves include the matter effect for the DUNE baseline.}
\label{fig:cpmat}
\end{figure}

{\color{black} We next investigate the impact of the matter potential on the sensitivity to $\delta_{\rm CP}$, as shown in Fig.~\ref{fig:cpmat}. For a fixed baseline of $L = 1300~\mathrm{km}$, the matter effect leads to a substantial enhancement of the QFI compared with the vacuum case, with the enhancement being particularly pronounced around $E \sim 2$--$3~\mathrm{GeV}$. In contrast, the corresponding modification of the transition probability $P(\nu_\mu \to \nu_e)$ is less pronounced than the enhancement observed in the QFI. This demonstrates that the matter potential can significantly modify the parameter sensitivity encoded in the quantum state even when its effect on the transition probability itself is less pronounced. The enhanced QFI therefore indicates an increased intrinsic sensitivity to $\delta_{\rm CP}$ in the matter environment relevant to DUNE.

An additional feature is the shift in the energy dependence induced by the matter potential. The maximum of the QFI moves from $E \simeq 2.6~\mathrm{GeV}$ in vacuum to $E \simeq 2.3~\mathrm{GeV}$ in matter, while the maximum of $P(\nu_\mu \to \nu_e)$ shifts from approximately $E \simeq 2.2~\mathrm{GeV}$ to $E \simeq 2.1~\mathrm{GeV}$. Thus, the matter potential modifies not only the magnitude of the sensitivity but also the energy at which the quantum state carries maximal information about $\delta_{\rm CP}$. The different locations of the QFI and probability maxima further emphasize that a large transition probability does not necessarily correspond to maximal parameter sensitivity. This highlights the complementary information provided by the QFI in identifying the optimal energy region for $\delta_{\rm CP}$ estimation.}

\begin{figure}[!htb]
\centering
\includegraphics[width=0.48\textwidth]{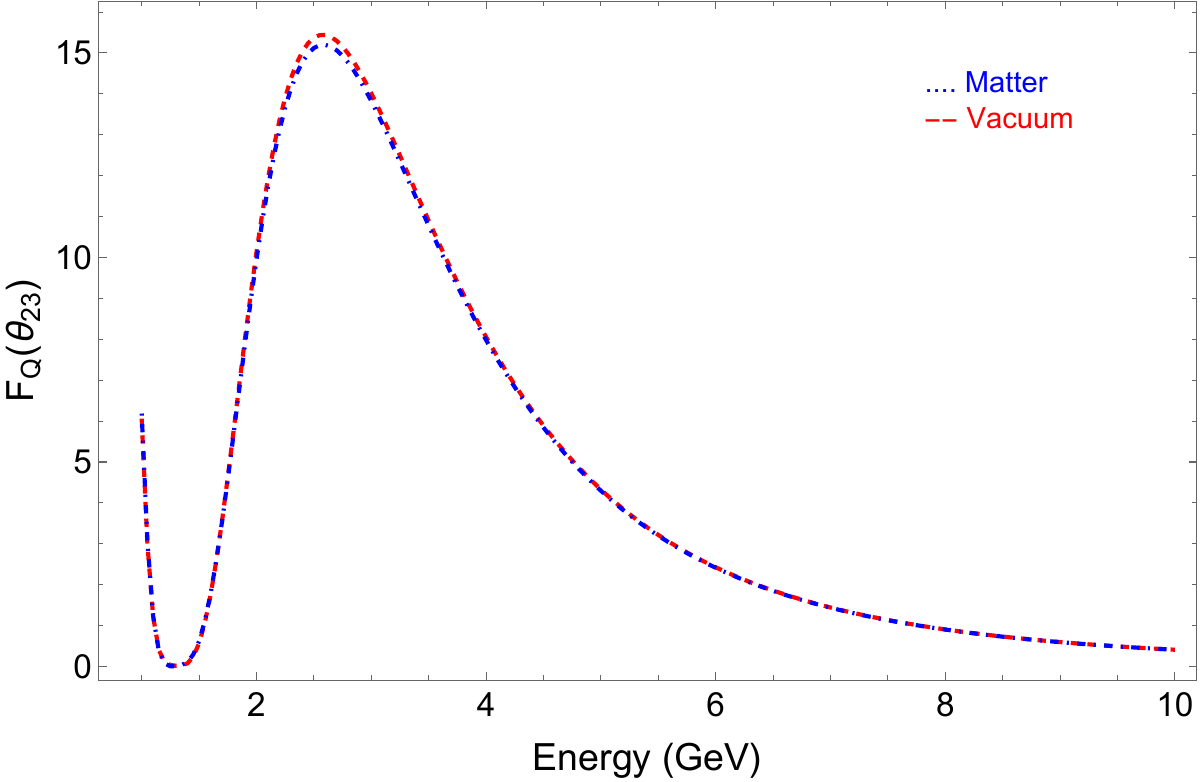}
\includegraphics[width=0.48\textwidth]{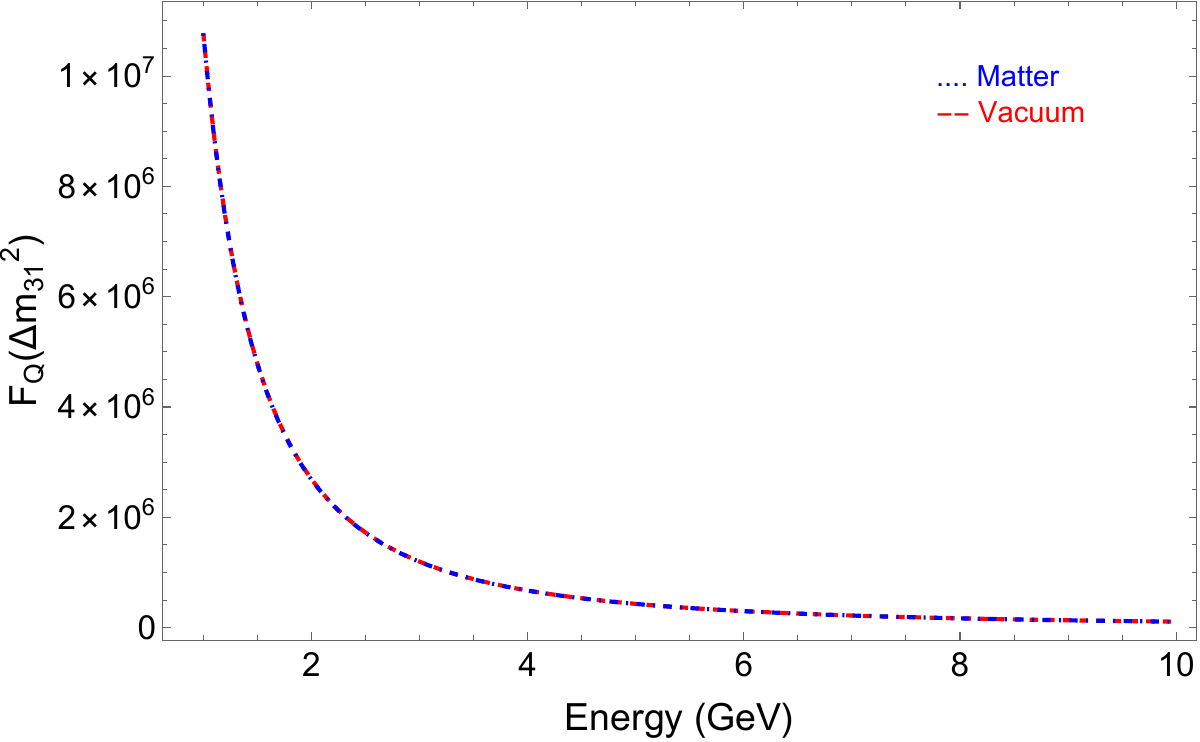}
\caption{QFI $F_Q(\theta_{23})$ and $F_Q(\Delta m_{31}^2)$ as functions of neutrino energy $E$ for a fixed baseline of $L = 1300~\mathrm{km}$. The red dashed curves represent vacuum propagation, while the blue dotted curves include the matter effect for the DUNE baseline. The close overlap of the vacuum and matter curves indicates that matter effects have a negligible impact on the QFI for these parameters.}
\label{fig:delthetamat}
\end{figure}

{\color{black} We finally examine the impact of matter effects on the QFI associated with $\theta_{23}$ and $\Delta m_{31}^2$, as shown in Fig.~\ref{fig:delthetamat}. In contrast to the pronounced enhancement observed for $\delta_{\rm CP}$, the QFI for both $\theta_{23}$ and $\Delta m_{31}^2$ remains nearly unchanged when matter effects are included. The matter and vacuum curves closely overlap across the entire energy range considered, indicating that the sensitivity to these two atmospheric parameters is largely insensitive to the presence of the matter potential for the fixed DUNE baseline. This distinct behavior demonstrates that the impact of matter effects on quantum parameter sensitivity is strongly parameter dependent, with $\delta_{\rm CP}$ being considerably more affected than $\theta_{23}$ and $\Delta m_{31}^2$. Thus, while matter propagation plays an important role in enhancing the quantum sensitivity to $\delta_{\rm CP}$, the sensitivity to the atmospheric parameters $\theta_{23}$ and $\Delta m_{31}^2$ is predominantly governed by the vacuum oscillation dynamics.}

This study extends previous quantum estimation analyses of $\delta_{\mathrm{CP}}$~\cite{Ignoti:2025rxr} to include $\theta_{23}$ and $\Delta m_{31}^2$, providing a unified framework to characterize both the intrinsic quantum sensitivity and the information accessible through flavor measurements. While the QFI establishes the ultimate quantum bound on the precision of parameter estimation, the CFI quantifies the information extracted through the specific flavor-transition probabilities considered in this work. Their comparison therefore provides a measure of how efficiently the chosen measurement accesses the information encoded in the quantum state and identifies regions where measurement optimization could potentially improve the attainable sensitivity. We further show that matter effects modify the quantum sensitivity in a parameter-dependent manner, producing a pronounced enhancement for $\delta_{\rm CP}$ while leaving the QFI for $\theta_{23}$ and $\Delta m_{31}^2$ largely unchanged. This framework can be naturally extended to multiparameter estimation, where correlations among oscillation parameters may influence the attainable precision and the compatibility of their simultaneous estimation.

\section{Conclusion}
\label{sec:con}

{\color{black}To summarize, we have investigated QFI as a fundamental measure of parameter sensitivity in three-flavor neutrino oscillations, providing a measurement-independent assessment of the ultimate precision limits imposed by quantum mechanics. Focusing on single-parameter estimation of the CP-violating phase $\delta_{\rm CP}$, the atmospheric mixing angle $\theta_{23}$, and the mass-squared difference $\Delta m_{31}^2$ in the $\nu_\mu \to \nu_e$ appearance channel. Using the NuFit-6.0 IC24 and IC19 benchmark parameter sets, we find a pronounced hierarchy in the intrinsic quantum sensitivity. The peak QFI values are approximately $F_Q(\delta_{\rm CP}) \sim 0.15$ and $F_Q(\theta_{23}) \sim 15$, while $F_Q(\Delta m_{31}^2)$ reaches values of order $10^7$ over the displayed $L/E$ range. Thus, the sensitivity $\Delta m_{31}^2$ is several orders of magnitude larger than that to the angular parameters. The QFI for $\delta_{\rm CP}$ and $\theta_{23}$ exhibits a bimodal structure around $L/E \sim 500$ and $1500~\mathrm{km/GeV}$, corresponding closely to the first and second oscillation maxima in the transition probability, whereas the QFI for $\Delta m_{31}^2$ increases strongly with $L/E$ over the range considered. These distinct behaviors reflect the different roles of the oscillation parameters in the evolution of the neutrino quantum state. Through the quantum Cram\'er--Rao bound, the QFI establishes the fundamental precision limit associated with the available quantum information. The sensitivity regions also encompass the operational $L/E$ regimes of current and proposed long-baseline experiments, including T2K, NO$\nu$A, DUNE, and ESS$\nu$SB.

We further compare the QFI with the CFI obtained from the flavor-transition probabilities. While the QFI provides the ultimate quantum bound on the information available for parameter estimation, the CFI quantifies the information accessible through the specific flavor-probability measurement considered here. For $\theta_{23}$ and $\Delta m_{31}^2$, the CFI follows the QFI relatively closely in important sensitivity regions, whereas a more pronounced separation is observed for $\delta_{\rm CP}$. This difference demonstrates that the considered flavor measurement does not uniformly extract all of the information encoded in the quantum state and indicates potential for improved parameter sensitivity through measurement optimization, without assuming that the QFI itself specifies an experimentally optimal measurement.

The inclusion of matter effects reveals a strong parameter dependence. For a fixed DUNE baseline of $1300~\mathrm{km}$, matter effects substantially enhance the QFI for $\delta_{\rm CP}$, particularly in the $E \sim 2$--$3~\mathrm{GeV}$ region, while producing only minor changes in the QFI for $\theta_{23}$ and $\Delta m_{31}^2$. This demonstrates that the matter potential can significantly modify the intrinsic quantum sensitivity to CP violation even when its effect on the corresponding transition probability is comparatively modest. The near overlap of the normal- and inverted-ordering results, together with the small differences between the IC24 and IC19 curves, further indicates that the principal quantum-sensitivity features are robust against the choice of mass ordering and the global-fit dataset.

Overall, the QFI--CFI comparison distinguishes the fundamental information available in the neutrino quantum state from the information accessible through a specific flavor measurement, while the matter-effect analysis demonstrates that this sensitivity is strongly parameter dependent. A natural extension of this work is a multiparameter treatment using the quantum Fisher information matrix (QFIM), incorporating correlations among $\delta_{\rm CP}$, $\theta_{23}$, $\Delta m_{31}^2$. Such an analysis would allow us to determine whether these parameters can be estimated independently or whether parameter correlations impose additional limitations on their simultaneous precision.}

\section*{Acknowledgments}
This work was supported by National Natural Science Foundation (Grant Nos. T2241005 and 12075059).

\appendix
\section{Experimental Parameters for Long-Baseline Neutrino Facilities}
\label{app:exp}

The representative $L/E$ values indicated in our results correspond to the characteristic kinematic regimes of accelerator-based long-baseline neutrino oscillation experiments. Table~\ref{tab:experiments} summarizes the baseline length $L$, peak neutrino energy $E_{\rm peak}$, and the resulting $L/E$ ratio for each facility.

\begin{table}[h]
\centering
\renewcommand{\arraystretch}{1.3}
\caption{Baseline configurations and characteristic $L/E$ values for long-baseline neutrino experiments.}
\label{tab:experiments}
\begin{tabular}{lccc}
\hline\hline
Experiment & $L$ (km) & $E_{\rm peak}$ (GeV) & $L/E$ (km/GeV) \\
\hline
MINOS~\cite{MINOS:2007ixr} & 735 & 3.0 & 245 \\
T2K~\cite{T2K:2025wet, Hu:2024qlx} & 295 & 0.6 & 490 \\
NO$\nu$A~\cite{Kalitkina:2025hbg} & 810 & 2.0 & 405 \\
DUNE~\cite{DUNE:2015lol, Perez-Molina:2026vex} & 1300 & 2.5 & 520 \\
ESS$\nu$SB~\cite{Giarnetti:2023pkz, Ghosh:2026cqk} & 360 & 0.25 & 1440 \\
\hline\hline
\end{tabular}
\end{table}
\section{Comparison between Normal and Inverted Ordering}
\label{app:qfi}
We also extend the single-parameter QFI analysis 
to compare the normal ordering~(NO) and inverted
ordering~(IO) scenarios. The purpose of this comparison is to assess the
dependence of the QFI on the neutrino mass ordering
and to examine the robustness of the intrinsic quantum sensitivity against
this discrete ambiguity. All results shown here correspond to the NuFit-6.0
IC24 benchmark dataset including S-K atmospheric data.
\begin{figure}[!htb]
\centering
\includegraphics[width=0.75\textwidth]{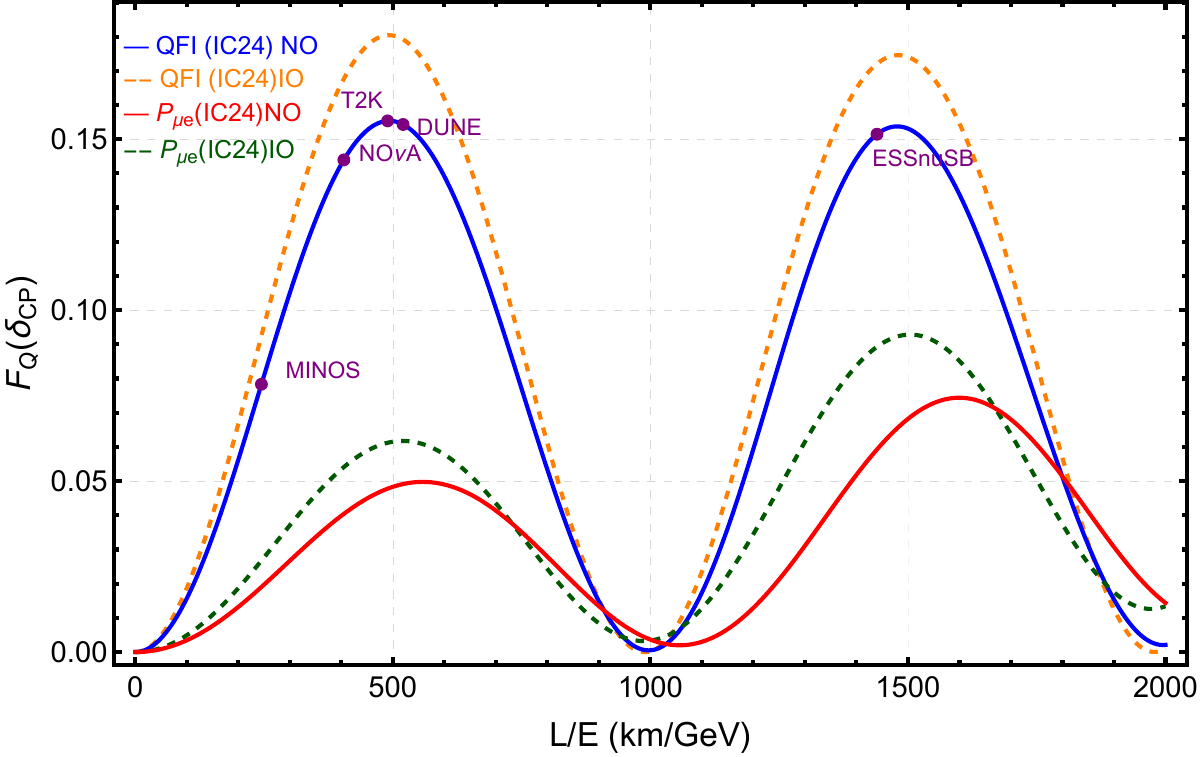}
\caption{QFI ${F}_Q(\delta_{\rm CP})$ and the
$\nu_\mu \to \nu_e$ oscillation probability as functions of $L/E$
for the NuFit-6.0 IC24 dataset including S-K atmospheric data.
The blue solid and orange dashed curves show the QFI for NO and IO, respectively.
The red solid and green dashed curves represent the corresponding
$\nu_\mu \to \nu_e$ transition probabilities for NO and IO.
}
\label{fig4}
\end{figure}

\begin{figure}[!htb]
\centering
\includegraphics[width=0.75\textwidth]{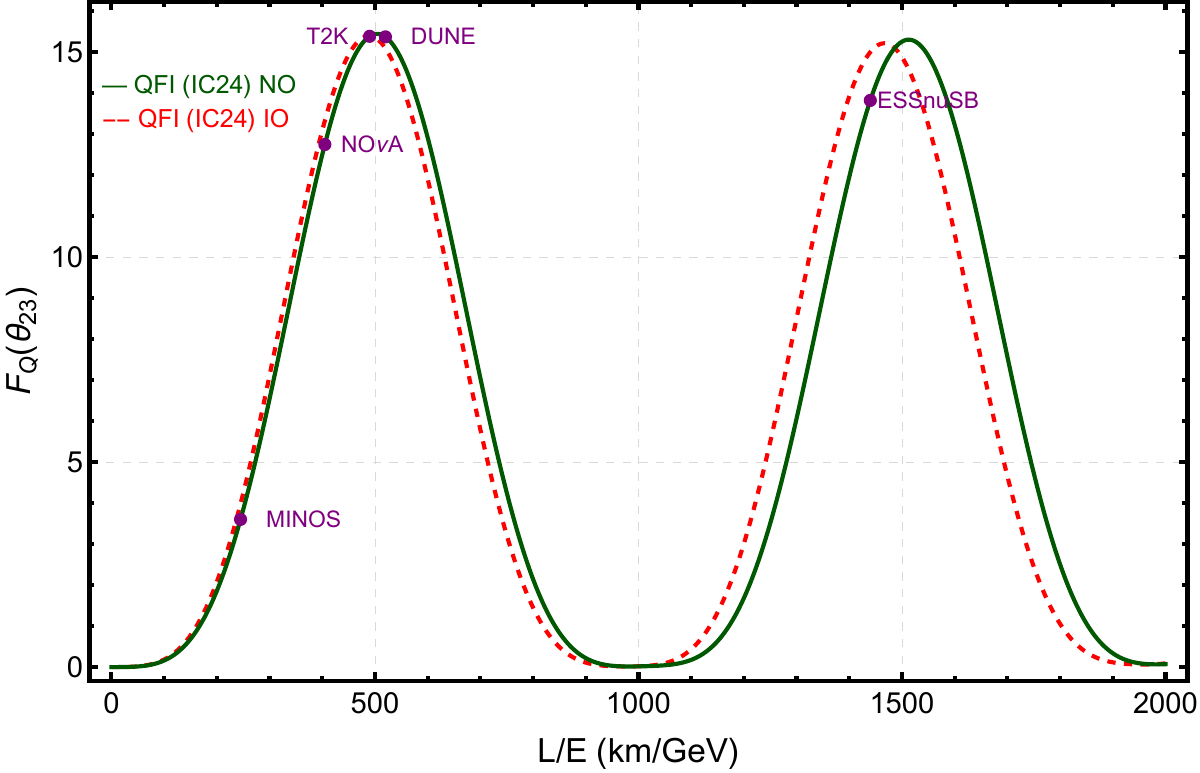}
\caption{
QFI ${F}_Q(\Theta_{\rm 23})$ as function of $L/E$
for the NuFit-6.0 IC24 dataset including S-K atmospheric data.
The green solid and red dashed curves show the QFI for NO
and IO, respectively.
}
\label{fig5}
\end{figure}

Figure~\ref{fig4} shows the QFI for the CP-violating phase
$\delta_{\mathrm{CP}}$ as a function of $L/E$ for both NO and IO. The QFI
peaks for NO and IO occur at identical $L/E$ values, indicating no
observable shift in peak positions. The peak magnitudes are comparable,
with a mild enhancement of the second peak for IO at the level of a few
percent, while the first peak remains unchanged.
\begin{figure}[!htb]
\centering
\includegraphics[width=0.75\textwidth]{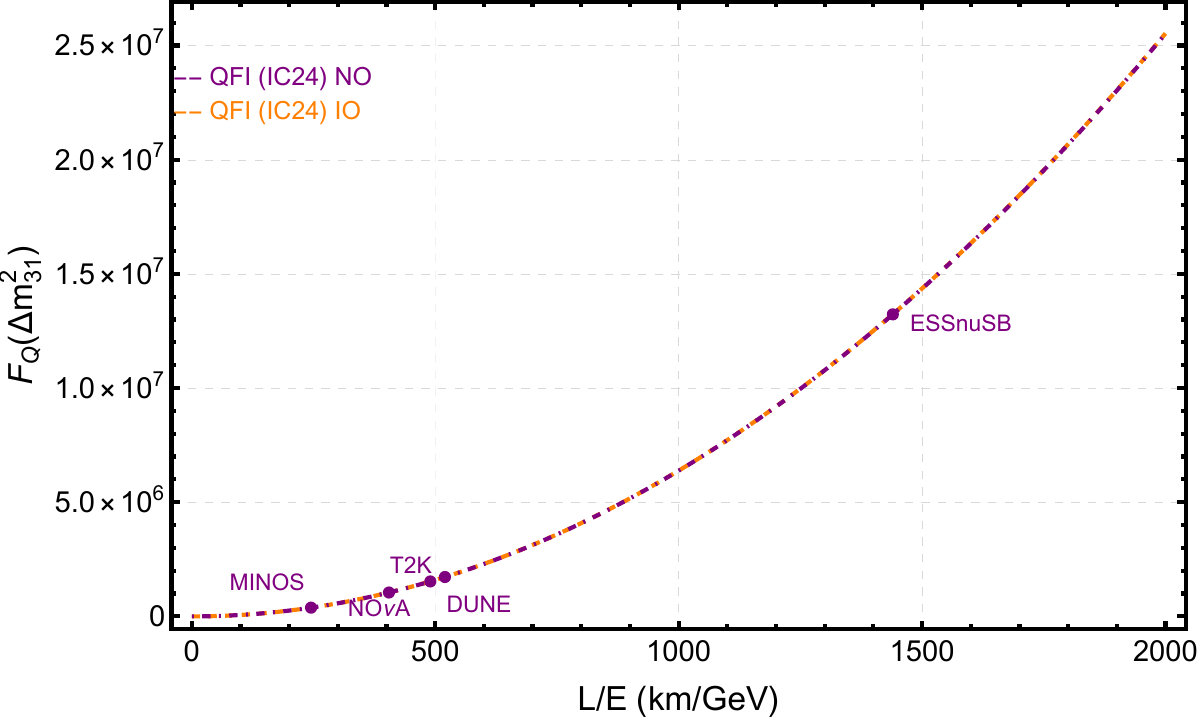}
\caption{
QFI ${F}_Q(\Theta_{\rm 23})$ as function of $L/E$
for the NuFit-6.0 IC24 dataset including S-K atmospheric data.
The green solid and red dashed curves show the QFI for NO
and IO, respectively.
}
\label{fig6}
\end{figure}

T2K and DUNE lie close to the first QFI maximum for both ordering,
indicating that these experiments operate near an optimal region for
probing $\delta_{\mathrm{CP}}$.
NO$\nu$A is located in an intermediate $L/E$ region between the first
maximum and the rising part of the QFI curve, while MINOS lies at
significantly lower $L/E$, where the QFI is comparatively small.
ESS$\nu$SB is positioned near the second oscillation maximum, which
corresponds to an enhanced QFI region for normal ordering and remains
close to the secondary peak for inverted ordering.

For $\theta_{23}$~(see Fig.~\ref{fig5}), the QFI peak heights are nearly identical for NO and IO, whereas one peak exhibit a noticeable shift in their $L/E$
position, constituting the dominant ordering-dependent effect for this
parameter. A similar behavior to that for $\delta_{\mathrm{CP}}$ is observed for the representative $L/E$ values of long-baseline experiments, with T2K and DUNE near the first QFI maximum,ESS$\nu$SB near the second maximum of IO, NO$\nu$A in an intermediate region, and MINOS at lower $L/E$.

{\color{black}In contrast, for $\Delta m_{31}^2$ (see Fig.~\ref{fig6}), the QFI curves for NO and IO remain essentially indistinguishable across the entire $L/E$ range, with no appreciable difference in either their magnitude or overall behavior. The QFI increases rapidly with $L/E$, reaching its largest value within the displayed range at $L/E \sim 2000~\mathrm{km/GeV}$. The representative $L/E$ values of MINOS, T2K, NO$\nu$A, DUNE, and ESS$\nu$SB are indicated by the corresponding points. The benchmark configurations therefore sample progressively higher regions of intrinsic QFI sensitivity as $L/E$ increases, with ESS$\nu$SB located at substantially larger $L/E$ than the other experiments. This behavior highlights the strong dependence of the quantum sensitivity to $\Delta m_{31}^2$ on $L/E$, while its near-independence of the mass ordering suggests that this sensitivity is relatively robust with respect to the choice of NO or IO.}


\label{Bibliography}
\bibliographystyle{JHEP}
\bibliography{refer}

\end{document}